\documentstyle[12pt,aasms]{article}

\def\la{\mathrel{\mathpalette\fun <}}

\def\fun#1#2{\lower0.837ex\vbox{\baselineskip0ex\lineskip0.209ex
  \ialign{$\mathsurround=0ex#1\hfil##\hfil$\crcr#2\crcr\sim\crcr}}}

\def\msun{M_\odot}
\def\msunyr{M_\odot \ {\rm yr}^{-1}}
\def\sles{\lower2pt\hbox{$\buildrel {\scriptstyle <}
   \over {\scriptstyle\sim}$}}
 
\def\sgreat{\lower2pt\hbox{$\buildrel {\scriptstyle >}
   \over {\scriptstyle\sim}$}}

\def\la{\mathrel{\mathpalette\fun <}}

\begin{document}

\title{ On the $M_V$(peak) Versus Orbital Period Relation
    for Dwarf Nova Outbursts}

\author{  John K. Cannizzo$^1$}
\affil{e-mail: cannizzo@lheavx.gsfc.nasa.gov}
\affil{Goddard Space Flight Center}
\affil{NASA/GSFC/Laboratory for High Energy Astrophysics, 
Code 662, Greenbelt, MD 20771}
\authoraddr{NASA/GSFC/Laboratory for High Energy Astrophysics, 
Code 662, Greenbelt, MD 20771}

\vskip 1truein
\ {\it }$^1$ Universities Space Research Association
\vskip 1truein
\centerline{ to appear in the Astrophysical Journal
  1998,  January 20, vol. 493}
\received{1997 May 16      }
\accepted{1997 September 2 }

\begin{abstract}

We present computations for the accretion
disk limit cycle model 
    in an attempt  to explain 
    the empirical relation
for dwarf nova outbursts
between the peak visual absolute magnitude
and orbital period found by Warner.
For longer period systems
one sees intrinsically  brighter outbursts.
This is accounted for in the limit cycle
model by  the scaling with radius of the critical surface
density $\Sigma_{\rm max}$
        which triggers
the dwarf nova outbursts.
During the storage phase of the instability
the accretion disk mass must be less
than
  some maximum value, a value which
scales with radius and therefore
orbital period.
When the instability
is triggered and the accumulated
mass is redistributed
into a quasi steady state
disk in outburst,
the resultant peak optical flux
from the disk is a measure of the
total mass which was stored
in quiescence.
We compute light
curves for a range in
outer disk radius
(or equivalently, orbital period),
and find that our peak
values 
of $M_V$
are within $<1$ mag
of the observed relation
$M_V$(peak)$=5.64-0.259P_{\rm orbital}$(h)
for $2\la P_{\rm orbital}$(h)$ \la 8$.

\end{abstract}

\medskip
\medskip

{\it Subject Headings:}
        accretion,
            accretion disks $-$ instabilities $-$  
                      cataclysmic
    variables $-$ stars: individual (SS Cygni)

\section{ INTRODUCTION }

Dwarf novae are a subclass
of the cataclysmic variables $-$
interacting binary stars
with orbital periods of several hours
in which  a Roche lobe filling
K or M dwarf
secondary transfers matter 
at a rate ${\dot M}_T$
into an accretion disk surrounding
a white dwarf (WD) primary.
Dwarf novae are characterized
observationally by having
outbursts
of several magnitudes
which recur on time scales
of days to decades and which can
last
    from $\sim1$ day to several
weeks (Warner 1995).
The accretion disk limit cycle
model has been developed
and refined to account for the outbursts
(Meyer \& Meyer-Hofmeister 1981;
for recent reviews see Cannizzo 1993a and Osaki 1996).
In this model,
material is stored up
at large radii
     in a relatively inviscid
accretion disk as neutral gas,
and then dumped onto the central
star as ionized gas when a certain
critical surface density
is achieved somewhere in the disk.

Warner (1987) presented a thorough
systematic study of the
dwarf novae. 
He noted several interesting
correlations between various
attributes associated
with the outbursts.
Of particular interest
is his finding of a relation between
the peak absolute magnitude
of dwarf nova outbursts and their
orbital periods:
$M_V$(peak)$=5.64-0.259P_{\rm orbital}$(h).
In deriving this relation,
Warner made corrections for distance
and inclination for $\sim20$ systems
that were well enough studied to
have reliable values for $M_V$(peak) and
orbital period $P_{\rm orbital}$.
The distance determinations
were made primarily on the basis
of infrared fluxes of the secondary stars,
as explained in Warner (1987).

In this work we quantify
the theoretical relation between
$M_V$(peak) and $P_{\rm orbital}$
by running computations of our time
dependent code which  describes  the
accretion disk limit cycle model
for a range in values of $r_{\rm outer}$,
the outer radius of the accretion disk.
We compare our results to Warner's
empirical relation.

\vfil\eject

\section {BACKGROUND }

The physical basis for the limit cycle
lies within the vertical structure
of the accretion disk.
In particular,
the change in the functional form of the
opacity at $\sim10^4$ K
coincident with the transition
from neutral to ionized gas
leads to 
        a hysteresis relation between
the effective temperature $T_{\rm eff}$
and surface density $\Sigma$
        in vertical structure
computations carried out at a given annulus.
This hysteresis leads to local
maxima 
$\Sigma_{\rm max}$
 and minima
$\Sigma_{\rm min}$
 in $\Sigma$
when the locus of solutions
is plotted as $T_{\rm eff}$ vs. $\Sigma$.
The resultant series of steady state
solutions resembles an "S".
Instability analyses of the
equilibrium  vertical structure
solutions and their linearly
perturbed states
show  that the upper and lower branches
of the "S" curve, those
portions where $d\log T_{\rm eff}/d\log\Sigma > 0$,
are viscously and thermally stable
and can accommodate physically
attainable states during quiescence and outburst.
The "middle" part of the "S"
where $d\log T_{\rm eff}/d\log\Sigma < 0$
is unstable and physically unattainable.
    During quiescence material accumulates
    at large radii in the disk
    because the viscosity is so low
    that there is little inward transport
of the gas.
    The disk is far from steady state.
When $\Sigma > \Sigma_{\rm max}$
somewhere in the disk, a heating
instability is triggered which begins an outburst;
and when  $\Sigma < \Sigma_{\rm min}$,
a cooling transition front is started
which controls the decay of the outburst
from maximum light.
     As the disk goes from quiescence
     to outburst,  the matter in the disk
     is redistributed from large to small
     radii, and the local ${\dot M}(r)$
  profile becomes  close to steady state,
  except for a strong outflow at large radii.
The local maximum in $\Sigma$ determined
from the vertical structure computations
is 
\begin{equation}
 \Sigma_{\rm max} = 11.4 \ 
{\rm g} \  {\rm cm}^{-2} \
 r_{10}^{1.05} \ 
 m_1^{-0.35}  \
 \alpha_{\rm cold}^{-0.86},
\end{equation}
where $r_{10}$ is the radius in units
of $10^{10}$ cm,
$m_1$ is the WD primary mass in solar units $M_1/1\msun$,
and $\alpha_{\rm cold}$
is the value of the Shakura \& Sunyaev (1973) viscosity
parameter on the lower (or neutral) stable branch
of the $S-$ curve (Cannizzo \& Wheeler 1984).
    (To avoid confusion,
   we use the subscript ``max'' to 
   refer to properties of the accretion
  disk associated with $\Sigma_{\rm max}$,
  and ``peak'' to refer to either the peak
  flux or peak accretion rate during  an outburst.)
The values for $\Sigma_{\rm max}$
found by other workers are similar,
reflecting slight differences
in the equations of state, opacities,
or treatments of the boundary conditions.
The largest uncertainties
may turn out to be due to the treatment
of convection in the vertical structure.
     Within   the confines
of standard mixing length theory
which is used in the vertical
structure computations,
     convection may in fact
be less important in dwarf novae
accretion disks than was previously 
thought (Gu \& Vishniac 1998).
Nevertheless,
even if convection is set to zero,
one still obtains
the hysteresis between
$T_{\rm eff}$ and $\Sigma$
(see Fig. 4 of 
Ludwig et al 1994).

Cannizzo (1993b, hereafter C93b) commented briefly on
Warner's finding of the empirical relation
between the peak flux level in a dwarf
nova outburst and the orbital period:
$M_V$(peak)$=5.64-0.259P_{\rm orbital}$(h).
C93b used the following
argument to derive a scaling law
for
the
rate of accretion onto the WD during the
peak of a dwarf nova outburst.
One can express the mass
of the disk 
at the end of the quiescence
interval during which time  material
   accumulates in the disk
as $f M_{\rm max}$,
where
$ M_{\rm max} = \int 2\pi r dr \Sigma_{\rm max}$
is the  ``maximum mass'' that the disk
could  have reached in quiescence
if the disk were filled
up to the level $\Sigma_{\rm max}$
at every radius.
Once the outburst has begun
and progressed for a short
time,
    the surface density profile
    adjusts  from one which was
     non-steady state in quiescence to one which
is in quasi-steady state in outburst.
This basically involves a sloshing
around of the gas from large to small
radii, so that $\Sigma_{\rm quiescence}\propto r$
and  $\Sigma_{\rm outburst}\propto r^{-1}$.
If this time interval is short,
one may equate the mass of the disk
at the end of the quiescence interval
to the mass of the disk at the
beginning of the outburst.
Therefore if one   integrates a
           Shakura-Sunyaev scaling for
$\Sigma_{\rm outburst}=$
    $\Sigma(r,\alpha,{\dot M})$ over the disk
as was done for the quiescent state by
taking $ M_{\rm outburst} = \int 2\pi r dr \Sigma_{\rm outburst}$, 
  sets this equal
to $f M_{\rm max}$, and then inverts
this expression to obtain ${\dot M}$,
      one
     derives an approximation
  for the peak ${\dot M}$ value
in the disk during outburst 
\begin{equation}
{\dot M}_{\rm peak} = 1.1\times 10^{-8} \ 
\msunyr \ 
\left(\alpha_{\rm hot}\over 0.1\right)^{1.14}
\left(\alpha_{\rm cold}\over 0.02\right)^{-1.23}
\left(r_{\rm outer}\over 4\times 10^{10} \ {\rm cm}\right)^{2.57}
\left(f\over 0.4\right)^{1.43},
\end{equation}
(C93b),
  where $\alpha_{\rm hot}$
is the alpha value on the upper   
         (or ionized)
stable branch, and
$r_{\rm outer}$ is the outer radius of the accretion disk
(which is set by the orbital period).
The values of the parameters
entering into eqn. (2) have been scaled to
the values which C93b found to be
relevant for SS Cygni,
a dwarf nova with $P_{\rm orbital}=$ 6.6 h.
From Kepler's law $P_{\rm orbital}^2\propto a^3$,
where $a$ is the orbital separation.
So if $r_{\rm outer}\propto a$
   we basically have ${\dot M}_{\rm peak}$
scaling as $P_{\rm orbital}^{1.7}$,
assuming the the other parameters
in eqn. (2) do not vary with orbital period.
If the visual flux were to scale
linearly with ${\dot M}$ in the disk,
then this would imply $M_V$(peak)
$\propto -0.68 \log P_{\rm orbital}$,
a different functionality than that observed.
One additional consideration which
comes into play for dwarf novae
at increasingly longer orbital periods
     is that
$r_{\rm outer}$ varies
   nonlinearly with $a$
in the regime where the secondary 
star transitions from being less
massive
than the primary star to
being of 
comparable mass.

Warner (1995, see  his Fig. 3.10)
     over-plotted eqn. (2) with
the data taken from Warner (1987)
to show that the analytical
expression does a reasonable
job of characterizing
the observations.
The analytical expression for
${\dot M}_{\rm peak}$ derived above
is only useful in comparing  with 
observations,
however,
    if the variables appearing in eqn. (2)
do not vary appreciably with radius.
Of particular concern is the variable
$f$
which gives $M_{\rm outburst}/M_{\rm max}$
at the time of burst onset.
One might expect for $f$ to
vary significantly with orbital
period $P_{\rm orbital}$ or
secondary mass transfer rate
${\dot M}_T$, in which case
the dependence on $r_{\rm outer}$
which appears in  eqn. (2)
would be misleading.
Our aim in this work is to understand
the function $f$ by running
time dependent models
for a range in values of
$r_{\rm outer}$ and ${\dot M}_T$.

\vfil\eject

\section {THE MODELS }

We use the computer model described
in previous works (C93b, Cannizzo 1994, 
     Cannizzo et al. 1995,
  hereafter CCL).
This
is a one dimensional time dependent
numerical model which solves
explicitly for the evolution
of surface density and midplane
temperature in the accretion disk.
We carry this out by solving the mass
and energy conservation equations
written in cylindrical coordinates
and averaged over disk thickness.
The scalings which characterize
the steady state relationship
between $T_{\rm eff}$ and $\Sigma$
were taken from the vertical structure
calculations (Cannizzo \& Wheeler 1984,
C93b).
For the WD mass we adopt $M_1=1\msun$,
for the inner disk radius
we take
$r_{\rm inner}=5\times10^8$ cm,
and for the number of grid points $N=300$.
Our grid spacing 
is such that $\Delta r=\sqrt{r}$.
We compute the visual flux
as described in C93b
by assuming a face-on oriented disk
and taking the standard Johnson $V-$band
filters.
C93b presents many tests
of the numerical model to assess
    systematic effects.

For the $\alpha$ parameter
which characterizes
the strength of viscous
dissipation
and angular momentum transport
within the accretion disk
we utilize the form given in CCL,
$\alpha=\alpha_0(h/r)^n$,
where $n\simeq1.5$.
CCL quantified the use of
this form based on the observed
ubiquitous exponential decays
seen in soft X-ray transients and
dwarf novae, but
they were not the first to use it;
  it was introduced
by Meyer \& Meyer-Hofmeister (1984).
The normalization constant
$\alpha_0\simeq 50$ is based
on the magnitude of the 
$e-$folding time constant,
and the exponent $n$ determines
the functional form of the decay:
$n<1.5$ gives a faster-than-exponential
decay, and 
$n>1.5$ gives a slower-than-exponential
decay. 
   The
   reason for this
particular functional
form being the preferred one,
          based on a detailed examination
of the departure from steady state
conditions within the hot part of the
accretion disk, has been recently
provided by Vishniac \&  Wheeler (1996).
There is one shortcoming
in adopting this form for
dwarf novae which must be addressed.
This shortcoming
can be rectified if one
imposes an upper limit on $\alpha$,
which seems physically
reasonable.
Figure 7 of CCL
shows the failure
of the form $\alpha\propto(h/r)^{1.5}$
to effect a change in
the $e-$folding decay time
in an outburst
with outer disk radius,
or equivalently, orbital period.
For dwarf novae, however, it is well
known that there exists a 
linear relation
between the 
$e-$folding time constant
associated with the
decay of the dwarf nova outbursts
and the orbital period (Bailey 1975).
Therefore, the CCL
$\alpha$ value cannot
be applicable to
dwarf novae without modification.
Recent computations applying
the form $\alpha=\alpha_0(h/r)^{1.5}$
to integrations of the vertical structure
show that, if  $\alpha$ is  computed
self-consistently  within this
formalism for dwarf novae, 
       one derives
large values of $\alpha$
         in the ``outburst state'' $-$
meaning  the point on the
upper branch of the $\log T_{\rm eff} - \log\Sigma$
curve which lies  at the 
vertical  extrapolation of the
$\Sigma_{\rm max}$ value (Gu \& Vishniac  1998). 
In fact, the values are considerably larger
than can be tolerated within the
framework derived from the  theoretical constraint
imposed by the Bailey relation (Smak 1984, Cannizzo 1994)
$-$ this constraint being that $\alpha_{\rm hot}$
cannot exceed $\sim0.1-0.2$.
This limit will be reached, unfortunately, when 
$h/r \simeq 0.016 - 0.025$  if we strictly
take $\alpha=\alpha_0(h/r)^{1.5}$.
Since this value of $h/r$
         is exceeded  on the upper stable
branch of steady state solutions
for dwarf novae, we conclude that 
the physical mechanism responsible
for generating the viscous dissipation
and transporting angular momentum
in accretion disks
must
saturate to some $\alpha_{\rm limit}\sim 0.1-0.2$
so as to give the Bailey relation (Smak 1984).
Disks tend to be thinner in systems
with larger central masses,
generally speaking,
 so systems such as X-ray
novae (and AGN) do not run into this limit (Gu \& Vishniac 1998).
This explains why X-ray novae do not show a ``Bailey relation''
(see Fig. 3 of Tanaka \& Shibazaki 1996).
In light of these considerations, we
utilize the CCL $\alpha$ form in our computations,
    but in the limit
of large $\alpha$ we do not allow it to exceed
0.2.

The fact that we see
a system as a dwarf nova means
that the mass transfer rate
into the outer accretion disk
from the secondary star
cannot be too great or else
the disk would be in permanent outburst
(Smak 1983).
This reasoning
must apply   to the systems
used by Warner (1987)
in his compilation.
Shafter (1992)
considered the relative
frequency of dwarf novae
as a fraction of all cataclysmic
variables
in different period
bins longward of the $2-3$ h
period gap
in an attempt to understand
the variation of the rate of mass
transfer
${\dot M}_T$ 
from the secondary star
(feeding
into the outer accretion disk)
with orbital
period which we observe
in the dwarf novae.
Shafter
     concluded
that, by restricting
our attention solely to dwarf novae,
    we are probably not sampling
the long term
${\dot M}_T (P_{\rm orbital})$
value which characterizes
the secular evolution
of cataclysmic variables  as a whole
    (Kolb 1993).
Therefore we need not feel
reticent in adopting
an ${\dot M}_T (P_{\rm orbital})$ law
which serves only to
ensure
 ${\dot M}_T < {\dot M}_{\rm crit}$,
where
    ${\dot M}_{\rm crit}$
    is the value
of the secondary mass transfer
 ${\dot M}_T$ which
    must be exceeded for the entire disk to be stable
in the high state
of the "S" curve.
Stated another way, our concern
in this work is to adopt a law for
${\dot M}_T (P_{\rm orbital})$
which will produce dwarf nova
outbursts. This law is not related
to that which characterizes
the entire class of cataclysmic variables.
From a practical standpoint,
we find that
our computed values of $M_V$(peak)
are relatively
insensitive to the specific
${\dot M}_T$
at  a given orbital period.

We run computations for five values
of $r_{\rm outer}/10^{10}$ cm $-$
2, 3, 4, 5, and 6 $-$  while
at the same time
   scaling  the  value
  of  the rate of mass transfer ${\dot M}_T$
    feeding into  the outer disk
 so that ${\dot M}_T <$
${\dot M}_{\rm crit}=$
${\dot M}(\Sigma_{\rm min}) \simeq$
  $10^{16}$
g s$^{-1}$
$ (r_{\rm outer}/10^{10} \ {\rm cm})^{2.6} m_1^{-0.87}$
(Cannizzo \& Wheeler 1984).
For our canonical SS Cyg model,
we take $r_{\rm outer}=4\times10^{10}$ cm
and ${\dot M}_T=6.3\times 10^{16}$ g s$^{-1}$ (C93b, Cannizzo 1996).
At this value, the system is about a factor of 4
below ${\dot M}_{\rm crit}$.
Therefore, to determine      ${\dot M}_T$ for
other $r_{\rm outer}$ values
we scale the SS Cyg ${\dot M}_T$       by $r_{\rm outer}^{2.6}$.
We also ran a second series of computations 
with the normalization on ${\dot M}_T$
a factor of two smaller.

Figure 1 shows the computations
from the first series of models. In Figure 1a
we give the light curves,
and in Figure 1b we give the accretion
disk masses, shown in units of $M_{\rm max}$
for the relevant $r_{\rm outer}$ value.
        The assumption implicit in eqn. (2)
      that $f$ is constant with orbital period
appears to be  quite good: there is no noticeable drifting
of the disk mass relative to $M_{\rm max}$
as $r_{\rm outer}$ changes.
Figure 2 shows the computations
from the second series.
The results are 
similar to the first series.
     The equilibrium disk masses
      shift
to slightly lower values, as do the
         $M_V$(peak) values.

Figure 3 shows the results  taken
from the     previous figures
of $M_V$(peak) versus
orbital   period.
The results from the first series
are indicated by the hatched area
with hatched lines inclined at $\pm45^{\circ}$
with respect to the $x-$axis,
    whereas those from the
second series have hatched lines
inclined at $90^{\circ}$ and $180^{\circ}$.
The conversion between $r_{\rm outer}$
and  $P_{\rm orbital}$ was carried out using the
fitting formula  given in Eggleton (1983),
assuming a   secondary 
        mass $M_2 = 0.1\msun P_{\rm orbital}$(h).
    The hatched area 
accompanying each   series shows
the range allowed by taking
the disk to fill between 0.7 and 0.8
of the Roche lobe of the primary.

\section{ DISCUSSION AND CONCLUSION  }

We have run models
using the accretion disk limit  cycle model
for dwarf novae in an attempt
to understand Warner's relation
for dwarf nova  outbursts
$M_V$(peak)$= 5.64 - 0.259 P_{\rm orbital}$(h).
As noted in C93b,
the  observed  upper limit
is a natural  consequence of
the ``maximum mass'' of the accretion
disk that is  allowed by 
the critical surface density $\Sigma_{\rm max}$.
This value increases steeply with orbital period,
therefore the 
amount of fuel available  in a dwarf
         nova outburst also scales with
orbital period.
An important finding
is that, for a constant 
value of ${\dot M}_T/{\dot M}_{\rm crit}$,
the value $f$  which is the  ratio
of the accretion disk mass at outburst onset
to the ``maximum mass'' is relatively constant
with orbital period.
This gives some confidence in the analytical
estimate  given by C93b.
The fact that the theoretical
variation of $M_V$(peak)  with
orbital period is
flatter than might have been
expected from C93b's scaling
is due in part to
the variation of the
orbital period with outer disk
radius in the limit where the
mass of the secondary star
starts to  become
   comparable to the primary mass.
This effect contributes to the arc-like
shape of the shaded regions shown in Fig. 3.

The mean level of $M_V$(peak)
in our models exceeds
Warner's empirical line
by $\sim0.3-0.5$ mag over
most of the range shown.
It is probable
      that our method 
for computing the
      $V$ band flux
is too crude
to expect consistency
with observations at this level
of detail $-$
for instance we do not include
           limb darkening
in the models,
an effect mentioned by Warner (1987).
Also, we have utilized Planckian
flux distributions
for the disk.
Wade (1988) has shown that,
to varying degrees,
both Planckian distributions
and stellar or Kurucz type distributions
fail to
represent adequately  the  observations.
Although Wade found that the Planckian
distributions
can account for both
UV flux and UV
    color in a sample of nova-like
variables,
the failure of the model  in other respects
leads to the conclusion
that ``one cannot
rely on model fitting
to give the correct luminosities
or mass-transfer rates
(within an order of magnitude).''
Unless there are systematic
effects which depend    strongly on
orbital period, however,
     our computed
$M_V$(peak)$-P_{\rm orbital}$ slope
should have some physical relevance.
When one goes beyond this to 
   considering
     the normalization level of $M_V$(peak),
         it would seem that
a better flux model must be utilized.

It is also interesting to note
the relative insensitivity
of $M_V$(peak) to 
${\dot M}_T$
at a given orbital period.
This would imply that 
         the observed scatter
at a given orbital period
must be due  largely to
variations in distance
and inclination.
We varied the normalization
constant on ${\dot M}_T$
by a factor of two
between our two series in this work,
and found the resulting
$M_V$(peak)
values to differ by   $\la0.3$ mag.

The  soft X-ray transients
$-$ interacting binary stars
containing a neutron star
or black hole as the accreting
object $-$
also seem to obey a relation
between $M_V$(peak) and $P_{\rm orbital}$
for outbursts.
Van Paradijs \& McClintock (1994)
noted a relation 
between $M_V$ and
$(L_X/L_{\rm Edd})^{1/2}$
$P_{\rm orbital}^{2/3}$,
where $L_X/L_{\rm Edd}$
is the X-ray luminosity in units
of the Eddington luminosity.
Irradiation is a complicating
factor 
in these systems, and by comparing
the $M_V$ values
  between Figure 1 of Warner (1987)
and Figure 2 of van Paradijs \& McClintock
(1994),
     one can see that the 
X-ray binaries are $\sim1-7$ mag
brighter than the dwarf novae
at maximum light.
    Clearly most of the optical
  flux is reprocessed X-ray radiation
  coming from large radii
  in  the disk.
For the black hole X-ray binaries,
however,
  a large fraction of the difference
comes from 
     having larger disks
due to the larger orbital
separations, at a given orbital period
(Cannizzo 1998).
For the neutron star systems
(in which the primary mass is $\sim1\msun$
  as in dwarf novae),
the difference is entirely due to the irradiation.


We thank
the following people
for allowing us generous use
of CPU time on
 their DEC AXP
workstations:
Thomas Cline and Johnson Hays
  in the Laboratory
for High Energy Astrophysics at Goddard;
Clara Hughes, Ron Polidan, and George Sonneborn
in the Laboratory for Astronomy and Solar Physics
at Goddard; and
Laurence Taff,
Alex Storrs,
and Ben Zellner at the Space Telescope Science
Institute.
     JKC
 was supported 
through the visiting
scientist program under
 the Universities Space Research Association
(USRA contract NAS5-32484)
in the Laboratory for High Energy Astrophysics
at Goddard Space Flight Center.

\vfil\eject
\centerline{ FIGURE CAPTIONS }

Figure 1. (a) The light curves from the first series
for which ({\it top to bottom})
the values $(r_{\rm outer}/10^{10} \ {\rm cm},$
$ {\dot M}_T/10^{16} \ {\rm g} \ {\rm s}^{-1}) =$
(2, 1.04), (3, 2.98), (4, 6.30), (5, 11.3), and
(6, 18.1).
The dashed lines indicate
      the approximate absolute magnitude
of the secondary star 
   $M_V = 22.0 - 17.46 \log P_{\rm orbital}$(h)
(Patterson 1984).
The two values shown in each panel 
are consistent with a disk which fills 0.7 and 0.8,
respectively,
of the Roche lobe of the primary.
In the first two panels these lines are
off-scale toward fainter magnitudes;
in the third panel only the upper line is visible.

(b) The accretion disk masses corresponding
to the light curves in (a),
scaled to the ``maximum mass'' values
({\it top to bottom}) $M_{\rm max}=$
$2.713\times 10^{23}$ g,
$8.013\times 10^{23}$ g,
$1.727\times 10^{24}$ g,
$3.135\times 10^{24}$ g, and
$5.103\times 10^{24}$ g.

Figure 2. (a) The light curves from the second
series for which the ${\dot M}_T$ values
are a factor of two smaller than in Fig. 1.
The $r_{\rm outer}$ values are the same.

(b) The accretion disk masses corresponding
to the light curves in (a),
scaled to the same maximum mass values as in Fig. 1.

Figure 3. The peak absolute magnitudes
of the outbursts shown in Figs. 1a and  2a.
The upper hatched region is for the first
series and the lower hatched region
is for the second series.
The normalization on the ${\dot M}_T$
values used for the two series is different
by a factor of two, and yet
         the offset between the two regions
is quite small.
To  convert between  $r_{\rm outer}$
and
   $P_{\rm orbital}$,
        we use  the fitting formula
for the Roche radius given by Eggleton (1983),
assuming either 0.8 (left boundary of
each hatched region) or 0.7
(right boundary of each hatched region)
for the fraction of the Roche lobe 
which $r_{\rm outer}$ occupies.
We also take $M_2 = 0.1\msun P_{\rm orbital}$(h).
The dashed line indicates
$M_V$(peak)$ = 5.64-0.259 P_{\rm orbital}$(h)
(Warner 1987).
Warner quotes  an rms scatter about this relation
of $\pm0.23$ mag.

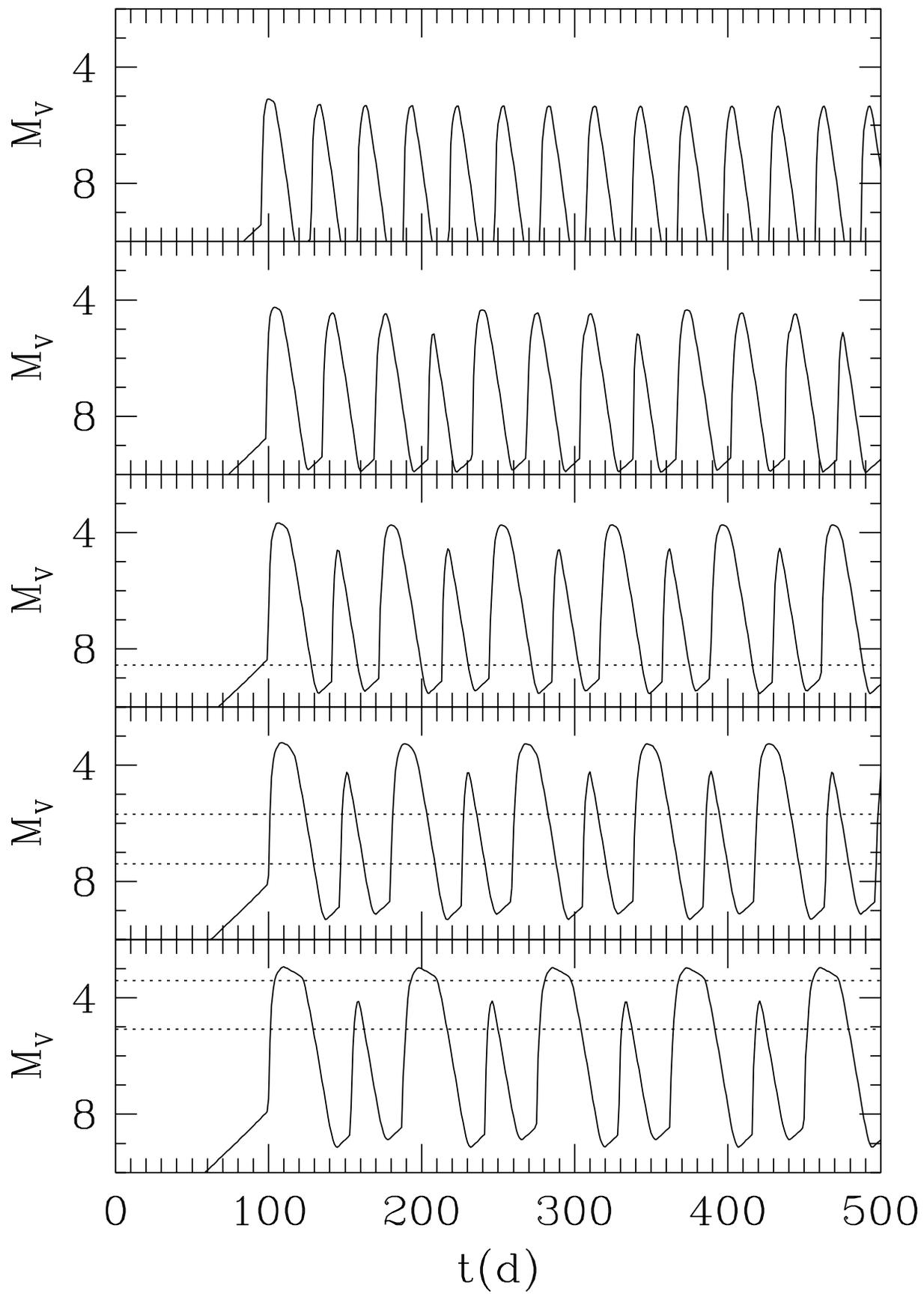
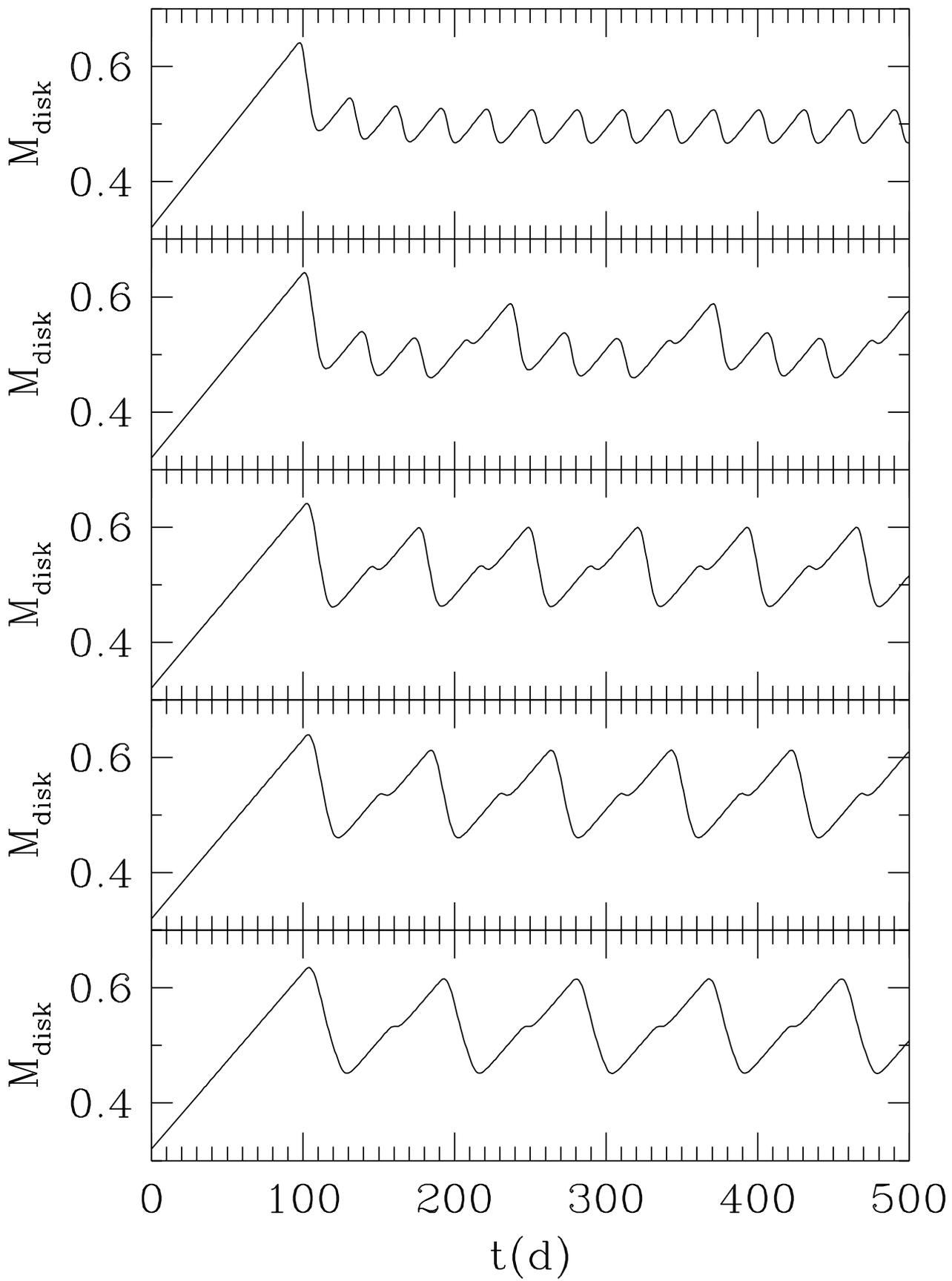
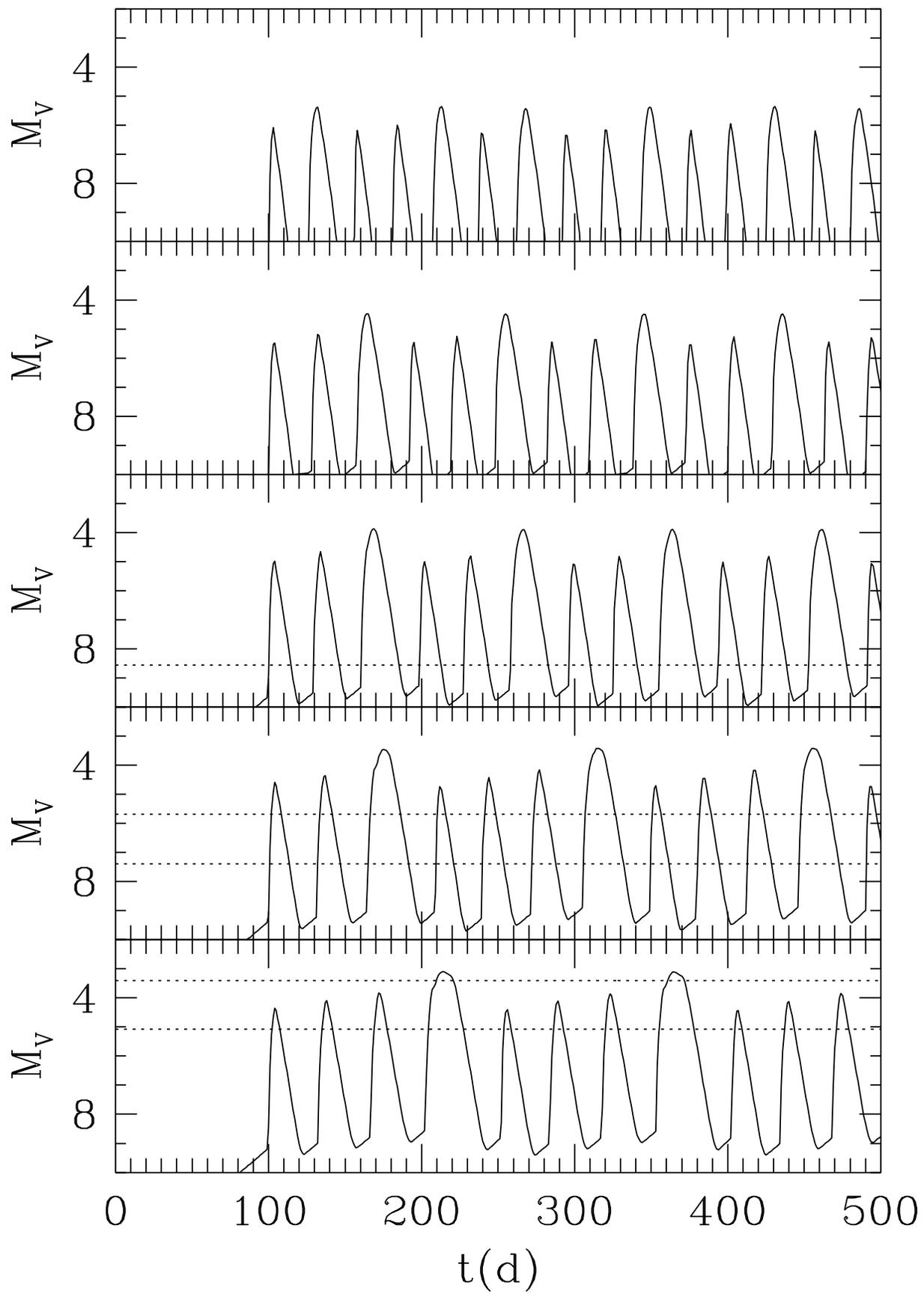
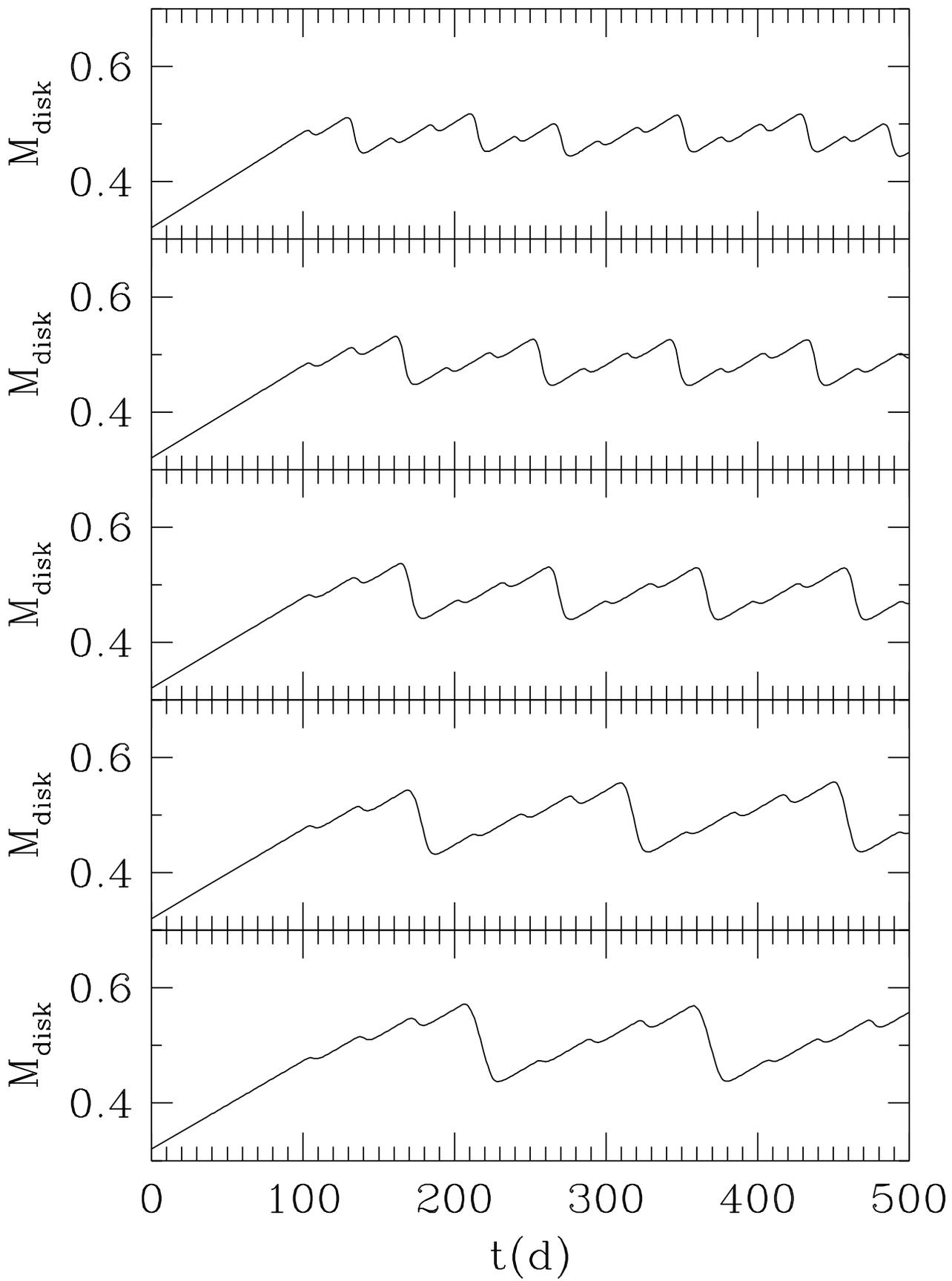
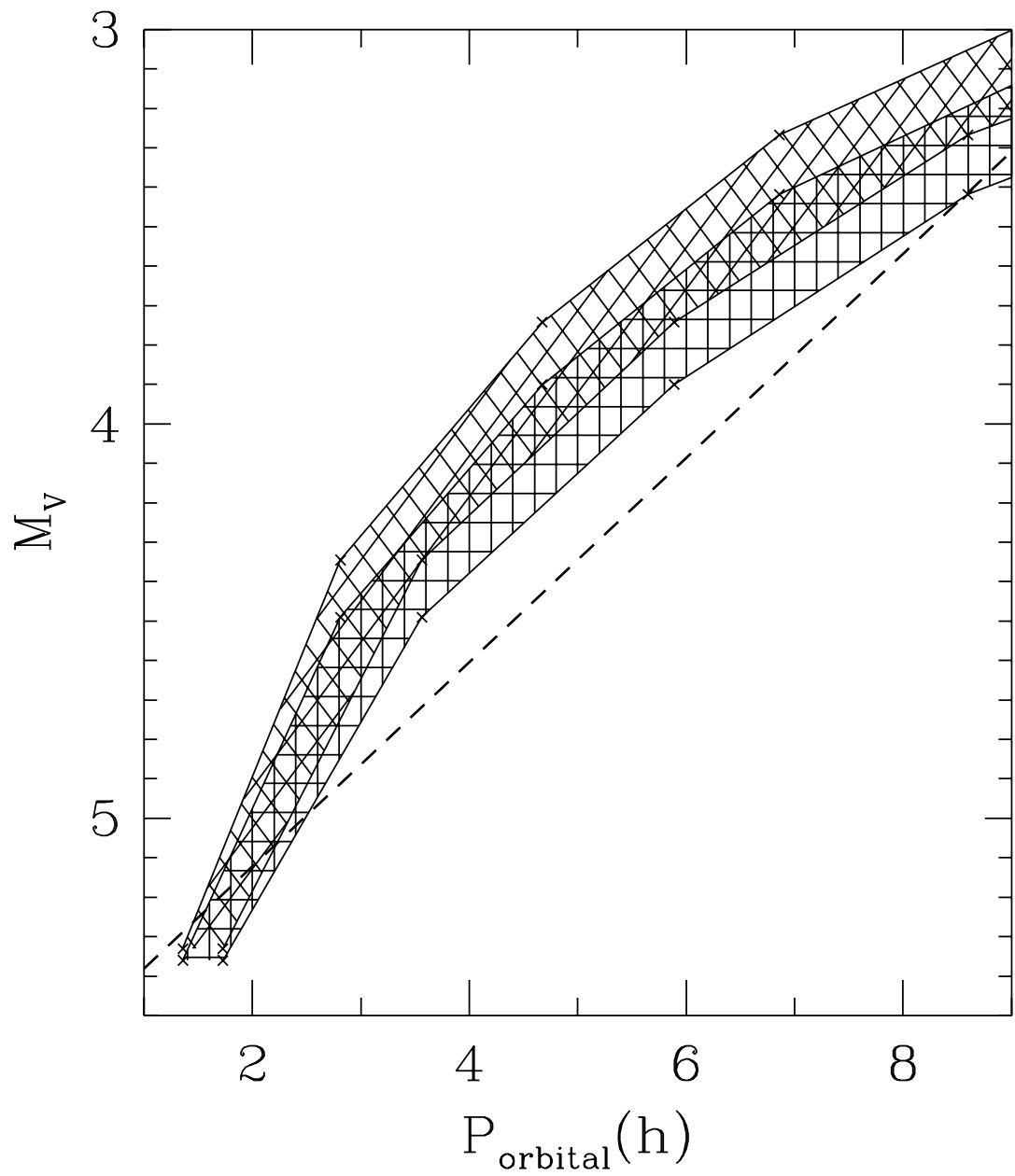
\end{document}